\documentclass[aps,prd,amssymb,showpacs,floatfix,twocolumn,superscriptaddress,nofootinbib]{revtex4-1}
\usepackage{amsmath,amsfonts,graphics,epsfig,amssymb,bm}
\usepackage{multirow}
\usepackage{enumerate}
\usepackage{graphicx}
\usepackage [latin1]{inputenc}
\usepackage[colorlinks, citecolor=blue, anchorcolor=blue, linkcolor=blue]{hyperref}

\usepackage[doipre={doi:~}]{uri}

\begin{document}

\title{Probing $\mu$eV ALPs with future LHAASO observation of AGN $\gamma$-ray spectra}
\author{Guangbo Long}
\author{Siyu Chen}
\author{Shuo Xu}
\author{Hong-Hao Zhang}
\email[Corresponding author. ]{zhh98@mail.sysu.edu.cn}
\affiliation{School of Physics, Sun Yat-Sen University, Guangzhou 510275, China}

\begin{center}
\begin{abstract}
 Axion-like particles (ALPs) are predicted in some well-motivated theories beyond the Standard Model. The TeV gamma-rays from active galactic nuclei (AGN) suffer attenuation by the pair production interactions with the cosmic background light (EBL/CMB) during its travel to the earth. The attenuation can be circumvented through photon-ALP conversions in the AGN and Galaxy magnetic-field, and a flux enhancement is expected to arise in the observed spectrum. In this work, we study the potential of the AGN gamma-ray spectrum for energy up to above 100\,TeV to probe ALP-parameter space at around $\mu$eV, where the coupling $g_{a\gamma}$ is so far relatively weakly constrained.
 We find the nearby and bright sources, Mrk\,501, IC\,310 and M\,87, are suitable for our objective. Assuming an intrinsic spectrum exponential cutoff energy, we extrapolate the observed spectra of these sources up to above 100\,TeV by the models with/without ALPs. For $g_{a\gamma}\gtrsim 2\times$$10^{-11} \rm GeV^{-1}$ with $m_{a}\lesssim0.5\,\mu$eV, the flux at around 100\,TeV predicted by the ALP model can be enhanced more than an order of magnitude than that from the standard absorption, and could be detected by LHAASO.
Our result is subject to the uncertainty from the intrinsic cutoff energy and the AGN lobe (or plume) magnetic-field. For an optimistic estimation, the constraint can be improved to $g_{a\gamma}\gtrsim 2\times$$10^{-11} \rm GeV^{-1}$ with $m_{a}\lesssim1\,\mu$eV. This require further observations on these sources by the forthcoming CTA, LHAASO, SWGO and so on.
\end{abstract}
\end{center}
\maketitle

\section{Introduction}
\label{sec:intro}
Several extensions of the Standard Model suggest the existence of very light pseudoscalar bosons called axion-like particles (ALPs) \cite{Svrcek2006,Jaeckel2010}. These spin-0 neutral particles are a sort of generalization of the axion, which was originally proposed to solve the strong CP problem naturally \cite{PQ1977,Weinberg1978,Wiczek1978}, and they are also a promising dark-matter candidate \cite{Preskill1983,Abbott1983,Dine1983,Marsh2011}.
One of their characteristics is their coupling to photons by $ g_{a\gamma} \mathbf{E}\cdot\mathbf{B} a$, with $g_{a\gamma}$ being the coupling strength, $\mathbf{E}$ the electric field of the photons, $\mathbf{B}$ an external magnetic field, and $a$ the ALP field strength \cite{Raffelt1988,Sikivie1983}. As a consequence, the phenomenon of photon-ALP mixing take place, and lead to photon-ALP oscillations (or conversions) \cite{Raffelt1988,Sikivie1983,Hochmuth2007,Angelis2008}.
To reach efficient conversions, they should take place above a critical energy given by \cite{Hooper2007,Angelis2007,Mirizzi2009}
\begin{equation}
E_{\rm crit}\sim38(\frac{m_{a}}{10^{-6}\rm eV})^{2}(\frac{10^{-5}\rm G}{B})(\frac{6.5\cdot
10^{-11}{\rm GeV}^{-1}}{g_{a\gamma }})\,\rm  TeV,\label{Ec}
\end{equation}
where $B$ is the homogeneous magnetic-field component transverse to the propagation direction. In contrast to axion, ALP mass $m_a$ is independent of $g_{a\gamma}$.
Around the critical energy, oscillatory features that depend on the configuration of magnetic field are expected to occur \cite{Wouters2012}.

 Many
laboratory experiments and astronomical observations are being carried out to search for ALPs via the effect mentioned above. The representative experiments are
photon-regenerated experiments, such as``Light
shining through a wall'' experiments ALPS \cite{ALPS2013}, solar ALPs experiments CAST \cite{CAST2011} and dark-matter haloscopes ADMX \cite{ADMX2006}.

Owing to the universal presence of magnetic fields along the line of sight to active galactic nuclei (AGN), photon-ALP oscillations can lead to distinctive signatures in AGN spectra \cite{Angelis2007,Simet2008,Hooper2007,Angelis2008,Meyer2014magnetic}. Thus, ALP-photon coupling can be detected through the observations of AGN (see, e.g. Refs \cite{Angelis2007,Conde2009,Angelis2011,Dominguez2011,Tavecchio2012}).

On the one hand, it is possible to detect the ALP-induced observational effects on the $\gamma$-rays transparency of the Universe \cite{Angelis2007,Simet2008,Conde2009,Angelis2011,Dominguez2011,Meyer2013,Meyer2014,Troitsky2016,Montanino2017,Buehler2020}. The very-high-energy (VHE, above 100~GeV) $\gamma$-rays from the extragalactic sources suffer attenuation by pair-production interactions with the background (extragalactic background light, EBL; or cosmic microwave background, CMB) photons during the propagation \cite{Nikishov1962,Hauser2001,HESS2006,Dwek2013,Costamante2013}. The attenuation increases with the distance to the source and the energy of the VHE photons \cite{Dwek2013}. If the photon-ALP conversions exist for a sufficiently large coupling, the emitted photons convert into ALPs and then these ALPs reconvert back into photons before arriving in the Earth, i.e. ALPs circumvent pair production. Thus, the opacity of the Universe for VHE gamma-rays is reduced and the observed flux is enhanced significantly (i.e. causing a hardening of the spectra above $E_{\rm crit}$, see e.g. Refs.~\cite{Angelis2007,Mirizzi2009,Dominguez2011,Angelis2013,Meyer2013,Troitsky2016,hardening2,Galanti2018,Galanti2019}). The range of the parameters where ALPs would increase the $\gamma$-ray transparency of the Universe (for 1.3 times the optical depth of Franceschini \emph{et al.} EBL model \cite{Franceschini2008}) is constrained from VHE $\gamma-$rays observations of blazar (AGN with jet closely aligned to the line of sight) \cite{Meyer2013}. Data from the Fermi-LAT observations of distant (redshift $z>$0.1) blazar limit $g_{a\gamma}< 10^{-11} \rm GeV^{-1}$ for $m_a<$3 neV assuming a value of the intergalactic magnetic field strength of 1 nG \cite{Buehler2020}.

On the other hand, taking seriously the irregularities of AGN gamma-ray spectra produced by the oscillations in cluster magnetic-field at energies around $E_{\rm crit}$, strong bounds on $g_{a\gamma}$ have been derived. In particular, for 0.4 neV$<m_a<$100 neV, the strongest bounds on $g_{a\gamma}$ are derived from the absence of irregularities in H.E.S.S. and Fermi-LAT observations as well as Fermi-LAT observations of AGN \cite{Hess2013alp,Fermi2016alp,Zhang2018alp,Li2020alp}. It is worth emphasizing that this method highly depend on the configuration of magnetic-fields adopted \cite{Libanov2020}.

So far, the coupling $g_{a\gamma}<6.6\times10^{-11}\,\rm GeV^{-1}$ for 0.2\,$\mu$eV$\lesssim m_a\lesssim2\,\mu$eV, containing viable
ALP dark matter parameter space (i.e. $g_{a\gamma}\lesssim2\times10^{-11}\,\rm GeV^{-1}$ for $m_a\sim\mu$eV) \cite{Arias2012}, almost have not been limited (see e.g. Figure.\,5 of Ref. \cite{Buehler2020}), although they are expected to be probed by future experiments (e.g. ALPS II \cite{ALPS2013}, IAXO \cite{IAXO2019}) or radio-astronomical observations (e.g. Refs \cite{Sigl2017,Edwards2020,Caputo2019,Ghosh2020}). According to Eq.\,(\ref{Ec}), effective oscillations for ALPs at this parameter range take place when the energy of photons is larger than $\sim$38\,TeV for the common value of AGN magnetic-field $B\sim10^{-5}$\,G \cite{Kohri2017,Fermi2016alp}. If considering more complicated and larger jet magnetic-field (e.g,. $10^{-4}$\,G, in kpc-scale jet), the energy $E_{\rm crit}$ for these ALP parameters can be less than 1\,TeV and the method of searching for the ALP-induced irregularities of blazar spectra may work \cite{Davies:2020uxn}. Nevertheless , the method based on the observational effects on the $\gamma$-rays transparency should be ineffective to search for these ALPs (even though the photons with energy $E_{\rm crit}\lesssim E \lesssim$38\,TeV can convert into ALPs in stronger jet magnetic-field), as the critical energy of reconversion from survival VHE photons to ALPs at Galaxy magnetic-field ($\sim1\,\mu$G) is above 300\,TeV. Therefore, photons with energy larger than 38\,TeV are required to be detected if probing these ALPs through a reduced opacity for AGN $\gamma-$rays. However, the highest energy photons detected from extragalactic sources are only about 20\,TeV so far \cite{Aharonian:1999vy,Abdalla:2019krx}.

 Thanks to the upcoming Large High Altitude Air Shower Observatory (LHAASO \cite{LHAASO2019}) with the ability to survey the TeV sky continuously, it is expected to reach sensitivities above 30\,TeV about 100 times higher than that of the current VHE instruments (e.g. H.E.S.S. \cite{HESSp}, MAGIC \cite{MAGICp}, VERITAS \cite{VERITASp}) \cite{LHAASO2019}. Furthermore, the conversions
in the intergalactic magnetic field (IGMF) can be neglected. With current upper limits on the IGMF strength of $\lesssim10^{-9}$\,G and on $g_{a\gamma} < 6.6\times10^{-11}\,\rm GeV^{-1}$ \cite{IGMF,CAST2011}, Eq.(\ref{Ec}) give that $E_{\rm crit}\lesssim$100\,TeV only for $m_a\lesssim16$\,neV.
Obviously, the unprecedented sensitivities of LHAASO to detect TeV AGN provide a good chance to probe $\mu$\,eV ALPs with the observational effects on the $\gamma$-rays transparency.

In this paper, we assume ALPs converted from the gamma-rays photons in the AGN's magnetic field travel unhindered through extragalactic space, and then these ALPs partially back-convert into photons in galactic magnetic field (GMF), see Fig.~\ref{fig:1}. We investigate the LHAASO sensitivity to detect the ALP-induced flux enhancement at the highest
energies by using the extrapolated observations of suitable AGNs for energy up to above 100\,TeV, and estimate the corresponding ALP-parameter space.

The paper is structured as follows. In section.\,\ref{sec:ALP} we give the formula for evaluating the photon survival probability along the line of sight. The sample selection is described and the
data analysis is introduced in section.\,\ref{sec:method} before presenting our results in section.\,\ref{sec:result}. We discuss our model assumptions in section.\,\ref{sec:discussion} and conclude in section.\,\ref{sec:conclusion}.

\section{Photon survival probability}
\label{sec:ALP}
\begin{figure}
\centering
\includegraphics[width=0.9\columnwidth]{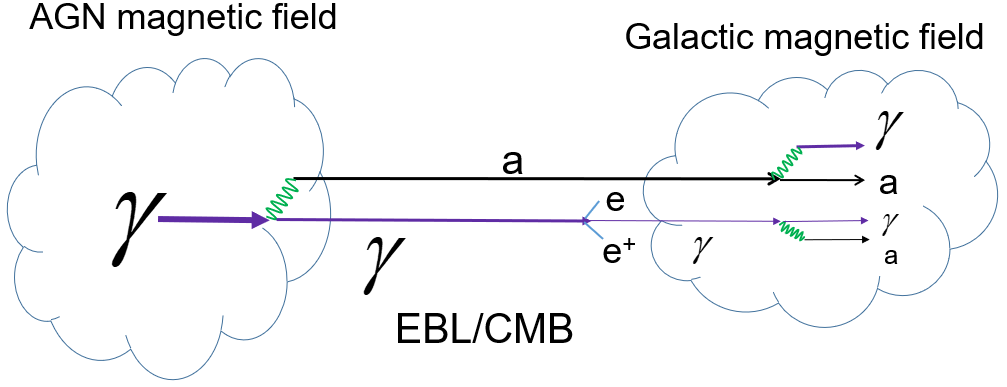}
\caption{\small{Cartoon of the formalism
  adopted in this article inspired by Ref.\,\cite{Conde2009}, where the TeV photon ($\gamma$, purple line) /ALP (a, black line) beam
  propagates from the AGN $\gamma$-ray source
  to the Earth. The interaction $\gamma+\gamma_{\rm EBL/CMB}\rightarrow \rm e^{\pm}$ takes place during the photon propagation. The photon-ALP conversions (green line) take place in the lobe magnetic-field around the gamma-ray emission region and GMF respectively, leading to an improvement of photon survival probability. There are two $\gamma\rightarrow\gamma$ channels: $\gamma\rightarrow\gamma (\rm e^{\pm})\rightarrow\gamma$; $\gamma\rightarrow a\rightarrow\gamma$, and the latter is dominant in this situation.}
  }
\label{fig:1}
\end{figure}

During the propagation from $\gamma$-ray emission regions to the Earth, we assume the emitted photons mix with ALPs in the AGN $B$-field and in the GMF respectively, and undergo the pair production with EBL/CMB in extragalactic space, shown in Fig.~\ref{fig:1}. The photon survival probability $P_{\gamma\rightarrow\gamma}$ will be calculated under these effects.
\subsection{Photon-ALP conversion}
\label{sec:conversion}
The probability of the conversion from an unpolarized photon to an ALP after passing through a homogeneous magnetic field $\mathbf{B}$ over a distance of length $r$, is expressed as \cite{Tavecchio2012,Angelis2013,Masaki2017}
\begin{equation}\label{probability}
P_{\gamma\rightarrow a}=\frac{1}{2}
\Big[\frac{g_{a \gamma}B}{\bigtriangleup_{\rm osc}(E)}\Big]^{2}{\rm sin}^{2}\Big[\frac{\bigtriangleup_{\rm osc}(E)r}{2}
\Big]
\end{equation}
 with
 $\bigtriangleup_{\rm osc}(E)$=$g_{a\gamma}B\sqrt{1+(\frac{E_{\rm crit}}{E}+\frac{E}{E_{\rm H}})^{2}}$, where $E_{\rm crit}$ is the critical energy defined in Eq.\,\ref{Ec} and $E_{\rm H}$ is derived from the Cotton-Mouton (CM) effect accounting for the photon one-loop vacuum polarization given by \cite{Tavecchio2012}
 \begin{equation}\label{E_H}
E_{\rm H}=
2.1\times10^{5}\big(\frac{10^{-5}\rm G}{B}\big)\big(\frac{g_{a\gamma}}{10^{-11}\rm GeV^{-1}}\big)\,\,\rm GeV.
\end{equation}
  The factor ``1/2" in Eq.\,\ref{probability} results from the average over the photon helicities \cite{Kohri2017}.

 $P_{a\rightarrow \gamma}$ (=$2P_{\gamma\rightarrow a}$) tends to be sizable and constant, when $E_{\rm crit}<E< E_{\rm H}$ and $1 \lesssim g_{a \gamma}Br/2$. For the case of the Galaxy, the latter can be expressed as
    \begin{equation}
1\lesssim(\frac{r}{10\,\rm kpc})(\frac{B}{1.23\,\mu \rm G})(\frac{g_{a\gamma}}{5\cdot
10^{-11}{\rm GeV}^{-1}}).\label{conversion_c}
\end{equation}
Here, we refer to Ref \cite{Han2017} for the values corresponding to $r$ and $B$.

\subsection{Magnetic field assumption}
\label{sec:magnetic}
The magnetic fields around the AGN gamma-ray source commonly include those in the jet, the radio lobes, and the host galaxy. The jet $B$ is believed to decrease as the distance to the central engine along the jet axis \cite{Tavecchio2015,Zheng2017,Meyer2014magnetic}. At the VHE emission region, the typical value of $B$ is in the interval of 0.1-5$\,$G \cite{Tavecchio2015,Zheng2017,Kang2014,Meyer2014magnetic,Xue2016,Sullivan2009}. The actual field strength in the radio lobes might be of order 1 to a few tens of $\mu$G in large \cite{Pudritz2012}, which depends strongly on the lobe size and power. The magnetic field in the host galactic is poorly known and its strength is approximately equal to $\mu$G with coherence lengths of the order of 0.1 to 0.2\,kpc \cite{Widrow2002,Meyer2014magnetic,IAXO2019}. A part of VHE $\gamma-$ray AGNs are located in galaxy clusters \cite{hardening2}, where the typical $B$ value is 5\,$\mu$G with coherence lengths of 10-100\,kpc \cite{IAXO2019,Widrow2002,Fermi2016alp}.

Our method takes advantage of the ALP-induced flux enhancement. It is more sensitive to the average size of magnetic fields than the complicated magnetic field configuration and detail with considerable uncertainty \cite{Meyer2013,Meyer2014magnetic,Fermi2016alp}. Furthermore, our study only give a estimation of LHAASO sensitivity to probe ALPs. Therefore, for simplicity, we assume the ``source" magnetic-field is homogeneous. According to Eq.\,\ref{Ec} and \ref{E_H}, the optimal size of magnetic field to probe the $\mu$eV ALPs within the preponderant energy range of LHAASO ($E>30$\,TeV) is $\sim$\,10\,$\mu$G. The magnetic-field at the VHE region is so high that the leading role played by the CM effect suppresses the photon-ALP conversion while the galaxy one is too weak, so the lobe B-field is expected to be responsible for the conversion site around the gamma-ray source. In this paper, the diffuse structures located at the terminal
of the jet including plumes are collectively referred to as ``lobes". If the $\gamma$-ray source locates in the cluster and thus the lobe should be inside the cluster too, we only consider the cluster B-field in the lobe and the lobe B-field should be stronger than without cluster.

  Due to the uncertainty of the lobe magnetic-field, we discuss the values of the B-field parameters including the strength $B_{\rm s}$ and region size $r_{\rm s}$ with two scenarios in our study. In the scenario of fiducial (slightly conservative) parameters, they are mainly inferred from observations for each source, see section.\,\ref{sec:method} and the fiducial parameters in Table.\,\ref{table:1} for details. In present, almost all of detected TeV AGN are Fanaroff-Riley I (FR\,I) radio galaxies and their aligned counterparts BL Lacs \cite{TEVCAT}. Due to the low luminosity for this type AGN and their lobe \cite{Urry1995}, the upper bound of $B_{\rm s}$ should be $\sim10\,\mu$G, and thus we take $B_{\rm s}$=$10\,\mu$G as the optimistic $B$ value of the lobe, see Table.\,\ref{table:1}. Besides, the AGN lobe is expected to be the candidate for Ultra-High-Energy Cosmic Rays sources, where the cosmic rays can be accelerated up to $\geq 10^{19}$\,eV and the ``Hillas condition" is satisfied. Then the size of conversion region satisfies $r_{\rm s}\geq10$\,kpc for $B_{\rm s}$=$10\,\mu$G. Coincidentally, the condition (\ref{conversion_c}) is also automatically satisfied for the value of $g_{11}$ (=$g_{a\gamma}\times10^{11}$\,$\rm GeV$) relevant to the present paper \cite{Kohri2017,Hooper2007}. Therefore, in the optimistic scenario, we assume $r_{\rm s}$=10\,kpc, where we take the lower limit of $r_{\rm s}$ given that the $B$ value is significantly large (even if larger $r_{\rm s}$ were considered, the change on the conversion probability could be small due to the strong conversion conditions (\ref{E_H}) and (\ref{conversion_c}) have already been satisfied). Note that the optimistic values of B-field were also applied in Refs.\,\cite{Kohri2017,Hooper2007} and suggested in Refs.\,\cite{IAXO2019} for the photon-ALP mixing in AGN jet or lobe.

 For the Galactic magnetic field (GMF), an average value of $B_{\rm GMF}$=1.23\,$\mu$G with scale 10\,kpc is assumed in our model\,\cite{Han2017}.

According to Eq.\,\ref{conversion_c}, the minimum coupling $g_{11}$ reaching the significant transformation is about 2.5 for the GMF. As the larger $B_{\rm s}$, the minimum $g_{11}$ corresponding to the conversion in the source can be low to 1. Analogously, the critical energy for the conversion in the GMF, $\sim$300\,TeV for $g_{11}\simeq6.5$, $m_{a}\simeq1\,\mu$eV, is higher than that (the critical energy) for the source. However, the
 CM effect can suppress the source conversion above 210\,TeV for $g_{11}=1$ due to the larger $B_{\rm s}$, as seen in Eq.\,\ref{E_H}, and it can be neglected in the case of GMF for the energy considered in this paper.

For more realistic models of jet magnetic-fields in connection with photon-ALP oscillations, the recent work in Ref.\,\cite{Davies:2020uxn} has investigated the effects of more complicated jet magnetic field configurations on the irregularities of GeV-10\,TeV spectrum. They find that this method can be used to search 1-1000\,neV ALP with $g_{a\gamma}\gtrsim5\times10^{-12}\,\rm GeV^{-1}$.

 We neglect the photon-photon dispersion effect on the propagation of TeV gamma rays. First, the dispersion contributed by CMB is dominant in the energy range of 100\,GeV to 1000\,TeV, even though radiation in the Galaxy or the source regions can far exceed the CMB \cite{Dobrynina2015}. Therefore, we only consider the CMB dispersion. Second, the CMB dispersion term in mixing matrix is $\Delta_{\rm{CMB}}\simeq 8 \times\left(\frac{E}{{100\,\rm TeV}}
\right)~{\rm Mpc}^{-1}$, and the photon-ALP mixing term $\Delta_{a\gamma}\simeq 15
\left(\frac{g_{a\gamma}}{10^{-11}\,\textrm{\rm GeV}^{-1}} \right)
\left(\frac{B}{10^{-6}\,\rm G}\right) {\rm Mpc}^{-1}$. When $\Delta_{\rm{CMB}}>2\,\Delta_{a\gamma}$, the dispersion effect is more important than the mixing, see e.g. Supplementary Material of Ref.\,\cite{Montanino2017}. Thus, the dispersion effect usually plays an important role in weaker magnetic field, e.g. the IGMF \cite{Galanti2019,Montanino2017}. But the photon-ALP mixing in IGMF is completely negligible in this work. For the Galaxy magnetic-field ($\sim 1\,\mu$G) and the coupling ($\gtrsim 10^{-11}\,\textrm{\rm GeV}^{-1}$) relevant to the present paper, the dispersion effect is important only when $E\gtrsim460$\,TeV.

\subsection{Photon survival probability}
\label{sec:probability}
We employ the EBL model of Ref. \cite{Gilmore2012} to account the VHE photon absorption onto the EBL. This recent EBL model has been tested repeatedly and is generally consistent with the VHE $\gamma$-ray observations
(e.g., Ref. \cite{Fermi2012,HESS2013,VERITAS2015,Biteau2015,MAGIC2016,Armstrong2017,Yuan2012,long2020}).
Furthermore, the infrared EBL intensity from this model is in the mid-level among several recent EBL models \cite{Dominguez2011EBL,Finke2010,Franceschini2008}, which are more inconsistent but basically match the direct measurements at infrared band \cite{HESS2017}. Hence, choosing this model to account the EBL optical depth is helpful to reduce EBL uncertainty. Above 140\,TeV, the CMB optical depth of TeV photons becomes dominant as its intensity is much stronger than the EBL's at the wavelength longer than 400\,$\mu$m.

After obtaining the EBL/CMB spectrum, we can further estimate the optical depth $\tau_{\gamma\gamma}(E, z)$ for the photon with energy $E$ from the source of redshift $z$ \cite{Gilmore2012,Gong2013}. Then, the photon survival probability on the whole path from the source to the earth can be derived
\begin{equation}\label{PSP}
P_{\gamma\rightarrow\gamma}=
  P_{\gamma\rightarrow \gamma}^{\rm S}{\rm exp}(-\tau_{\gamma\gamma})P_{\gamma\rightarrow\gamma}^{\rm G}+P_{\gamma\rightarrow a}^{\rm S}P_{a\rightarrow \gamma}^{\rm G},
\end{equation}
where $P_{\gamma\rightarrow a}^{\rm S}$ and $P_{a\rightarrow \gamma}^{\rm S}$ are the conversion probabilities from photons/ALPs to ALPs/photons in the source respectively. There is a relation $P_{\gamma\rightarrow \gamma}^{\rm S}=1-P_{\gamma\rightarrow a}^{\rm S}$; Similarly, the variables with a superscript ``G'' represent those for the GMF. The derivation of Eq.\,\ref{PSP} can be illustrated vividly in Fig.\,\ref{fig:1}: the first term corresponds to $\gamma\rightarrow\gamma (\rm e^{\pm})\rightarrow\gamma$ channel suffered from EBL/CMB absorption. the second is related to $\gamma\rightarrow a\rightarrow\gamma$ channel unaffected by the absorption.

\begin{figure}
\centering
\includegraphics[width=0.9\columnwidth]{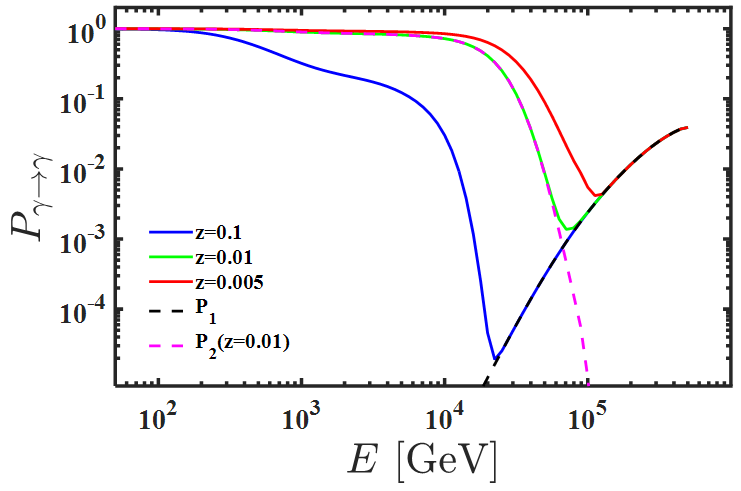}
\caption{\small{The photon survival probability $P_{\gamma\rightarrow\gamma}(E, z)$ on the whole path from the $\gamma-$ray source to the earth for the optimistic B-field parameters, where the ALP mass $m_a=1\,\mu$eV and coupling $g_{11}=3$. The meaning of each colored curve is annotated in the diagram. $P_{1}=P_{\gamma\rightarrow a}^{\rm S}P_{a\rightarrow \gamma}^{\rm G}$ and $P_{2}=P_{\gamma\rightarrow \gamma}^{\rm S}{\rm exp}(-\tau_{\gamma\gamma}(E, 0.01))P_{\gamma\rightarrow\gamma}^{\rm G}$, i.e., the second and first (redshift-independent) term in Eq.\,\ref{PSP}. They correspond to the channels of $\gamma\rightarrow a\rightarrow\gamma$ and $\gamma\rightarrow\gamma (\rm e^{\pm})\rightarrow\gamma$ shown in Fig.\,\ref{fig:1}, respectively.}}
\label{fig:pz}
\end{figure}
Fig.\,\ref{fig:pz} shows the change of $P_{\gamma\rightarrow\gamma}(E, z)$ with energy $E$ for different $z$, where the ALP mass $m_a=1\,\mu$eV and coupling $g_{11}=3$. $P_{2}=P_{\gamma\rightarrow \gamma}^{\rm S}{\rm exp}(-\tau_{\gamma\gamma}(E, 0.01))P_{\gamma\rightarrow\gamma}^{\rm G}$ and $P_{1}=P_{\gamma\rightarrow a}^{\rm S}P_{a\rightarrow \gamma}^{\rm G}$, i.e., the first and second term in Eq.\,\ref{PSP}. In the low energy region, the conversion is noneffective and $P_{\gamma\rightarrow\gamma}(E, z)$ is dominated by the absorption term $P_{2}\sim{\rm exp}(-\tau_{\gamma\gamma})$. As the energy turns to the higher region, ${\rm exp}(-\tau_{\gamma\gamma})\rightarrow0$, while the channel $\gamma\rightarrow a\rightarrow\gamma$ is getting ``wider'', and hence $P_{\gamma\rightarrow\gamma}\simeq P_{1}$, which is independent of $z$. As a consequence, the curves of $P_{\gamma\rightarrow\gamma}(E, z)$ for different $z$ at high energy region show v-shaped lines, and converge to $P_{1}$. When $E>E_{\rm crit}$, $P_{1}$ is getting closer and closer to its maximum, and when $E>E_{\rm H}\thickapprox$630\,TeV, the CM effect suppresses the source conversion. Thus the peak appears at the highest energy band.
\begin{table*}[t]
\centering
\caption{Magnetic fields and the best-fitting intrinsic spectra for the sources in the fiducial and optimistic cases. The values of the B-field parameters including the size of the homogeneous B-field $B_{\rm s}$ and region scale $r_{\rm s}$ respond to the AGN lobe. $\alpha$ is the photon index of the chosen intrinsic spectrum determined by the best fit. The BPLC spectrum has two spectral indices. In the fiducial case, the values of $B_{\rm s}$ are mainly inferred from references and $r_{\rm s}$ is conservatively assumed to 10\,kpc. The minimum cut-off energy $E_{\rm c}$ is determined by the best-fitting of $\psi_{0}$ (Eq.\ref{eq:model}, without ALP) to the observed spectrum, where the goodness of fit is measured by the minimum chi-square on every degree of freedom ($\chi^{2}$/d.o.f.). In the optimistic scenario, the values of $B_{\rm s}$ are roughly equal to the upper limit of the FR\,I and BL Lacs lobe. While the value of $r_{\rm s}$ is motivated from the assumption that the TeV sources are acceleration sites of high energy cosmic rays up to $10^{19}$\,eV at most \cite{Kohri2017,Hooper2007}, where $B_{\rm s}$ and $r_{\rm s}$ are satisfied the famous ``Hillas condition''. We take the cut-off energy $E_{\rm c}$=100\,TeV which is larger than the minimum one and the minimum $\chi^{2}$/d.o.f. remains unchanged. See text for further details.
\label{table:1} }
\renewcommand\tabcolsep{1pt}
\begin{ruledtabular}
\begin{tabular}{lcccccccccc}
   \multirow{2}{*}{Source} & & Fiducial & &Intrinsic spectrum
    & & & Optimistic& & Intrinsic spectrum&
     \\
   &$B_{\rm s}$\,($\mu$\,G) & $r_{\rm s}$\,(kpc)& $E_{\rm c}\,$(TeV)& $\alpha$\,(model) & $\chi^{2}$/d.o.f. \,\,\,\,\,\,\,\,\,\,\,\,\,& $B_{\rm s}$\,($\mu$\,G) & $r_{\rm s}$\,(kpc)& $E_{\rm c}\,$(TeV) & $\alpha$\,(model) &
    $\chi^{2}$/d.o.f. \\
\hline
M\,87& 5 & 10 & 90 &
 2.1\,(PLC) & 1.2 \,\,\,\,\,\,\,\,\,\,\,\,\,\,\,& 10 & 10 & 100 &2.1\,(PLC) & 1.2 \\
IC\,310& 2 & 10 & 40 &
 1.13,1.78\,(BPLC) & 0.18 \,\,\,\,\,\,\,\,\,\,\,\,\,\,\,& 10 & 10 & 100 &1.18,1.74\,(BPLC) & 0.18 \\
Mrk\,501& 2 & 10 & 70 &
 1.83,2.54\,(BPLC) & 1.4\,\,\,\,\,\,\,\,\,\,\,\,\,\,\,& 10 & 10 & 100 &1.83,2.58\,(BPLC) & 1.4 \\
\end{tabular}
\end{ruledtabular}
\end{table*}
Note that since we do not concern the spectral irregularities,
 we take the average value for the square of the sine function in Eq.~\ref{probability} when the phase is larger than 1 rad, e.g., according to Refs. \cite{Kohri2017,Hooper2007,Mirizzi2007}, smearing out the rapid-oscillatory features of the probability function. The average value is approximatively taken 2/3 rather than 1/2 to match the saturation-conversion probability ($P_{\gamma\rightarrow a}$) of 1/3, which corresponds to a more realistic scenario the beam propagates through many domains of randomly oriented magnetic fields with constant size $B$, for instance, in Refs. \cite{Mirizzi2007,Meyer2014}.

\begin{figure*}[t]
\centering
\includegraphics[width=0.45\textwidth]{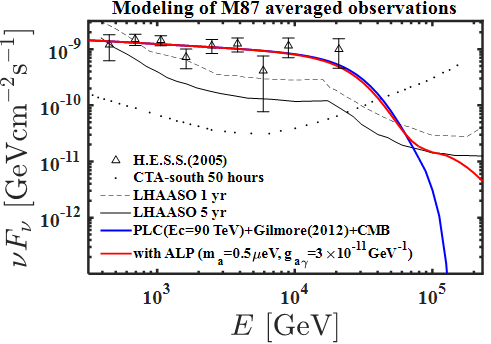}\,\,\,\,\,\,\,\,\,\,\,
\includegraphics[width=0.45\textwidth]{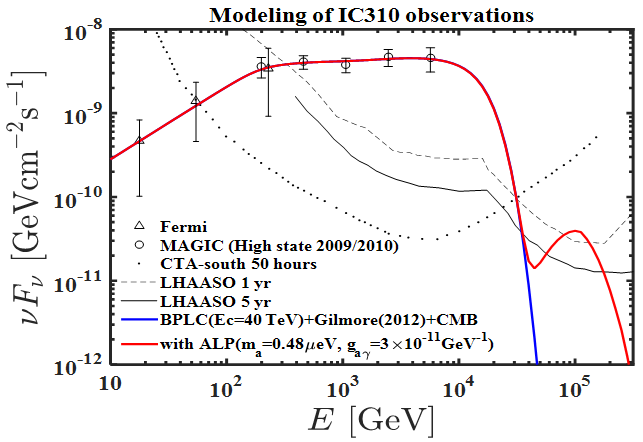}

\includegraphics[width=0.43\textwidth]{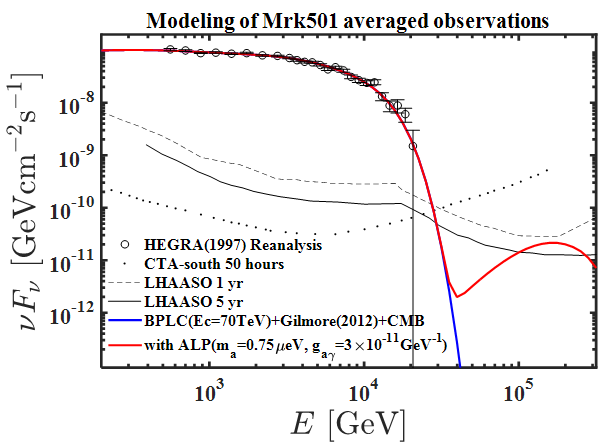}
\includegraphics[width=0.56\textwidth]{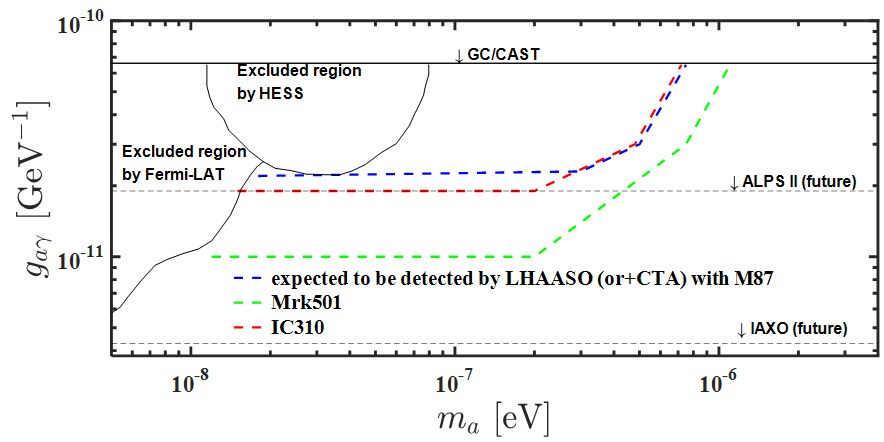}
\caption{\small{The results for the fiducial parameters in Table.\,\ref{table:1}. Top panel and left of the bottom panels: fitting and extrapolating the observations of M\,87, IC\,310 and Mrk\,501. The blue and red lines represents respectively the PLC or BPLC fit with the EBL (Gilmore)/CMB-absorption correction and that considering further the photon-ALP conversion and the CM effect. $E_{\rm c}$ is the minimum cut-off energy allowed by the best-fitting intrinsic spectrum. The meaning of other symbols are indicated in the legend. Right of the bottom panel: expected ALP limits, based on our model as well as the future LHAASO (or+CTA) observations of M\,87 (blue dash line), Mrk\,501 (green dash line) and IC\,310 (red dash line). For comparison, limits (black line) and 5\,$\sigma$ sensitivities of future experiments (black dashed line) are also shown.}}
\label{fig:constrain}
\end{figure*}

In the limit of saturated conversion $E_{\rm crit}\ll E \ll E_{\rm H}$ and $1 < g_{a \gamma}Br/2$, about
\begin{equation}\label{P_value}
 P_{\gamma\rightarrow a}^{\rm S}P_{a\rightarrow \gamma}^{\rm G}=\frac{1}{3}\times\frac{2}{3}
\end{equation}
 of the original photons survive through $\gamma\rightarrow a\rightarrow\gamma$ channel. Obviously, for the ALP-parameter value applied to Fig.\,\ref{fig:pz}, the condition of the saturated conversion does not match, for example $g_{a \gamma}B_{\rm GMF}r_{\rm GMF}/2 <1$.

\section{method}
\label{sec:method}

\subsection{Sample selection}
\label{sec:selection}
In our method, the VHE $\gamma-$ray observations are utilized to model the intrinsic spectrum of the emitted source. So far about 75 VHE AGN have been detected by the VHE instruments \cite{TEVCAT}. In principle, most of these sources may be used to search for the ALP-induced flux boost since it is independent of the redshift above $E_{\rm crit}$, as shown in Fig.\,\ref{fig:pz}. But we should acquire as many data below $E_{\rm crit}$ as possible, which are expected to be slightly affected by the ALPs, so that a more realistic spectrum at higher energy could be extrapolated by the observations together with the assumed model (see Eq.\,\ref{eq:model}). Hence, we preferentially consider the nearby sources whose $P_{\gamma\rightarrow\gamma}$ curve can show a ``shallow valley'' due to the relatively slight $\gamma\gamma$-absorption (see Fig.\,\ref{fig:pz}).

  Based on the study of Franceschini {\it et al.} \cite{Franceschini2019}, the adjacent sources of M\,87, IC\,310 and Mkn\,501 would likely be detected by LHAASO up to 75\,TeV, 50\,TeV, 25\,TeV respectively, when taking into account the standard EBL-absorption. Thus, we predict that they could provide more detectable data below $E_{\rm crit}$. In this paper, we will fit and extrapolate the spectral data of these sources to the highest VHE energies.

  \emph{M\,87}(Gal\,Long: 283.74, Gal\,Lat: 74.49; $z$=0.004)---a giant radio galaxy of FR\,I with kpc radio jet, located in the Virgo Cluster. It has
been detected by almost all the Imaging
Air Cherenkov telescopes (IACTs) \cite{M87MAGIC2019}. Strong and rapid flux variability in gamma-ray band was shown, but no significant spectral changes with a typical photon index of 2.2 \cite{M87MAGIC2019,M87VERITAS2011,M87HESS2006}. We adopt H.E.S.S. data taken during 21\,h of effective observation,
during the 2005 12.\,Feb.-15.\,May high state (see Fig.\,\ref{fig:constrain}).

From the results of minimum-pressure analysis for the magnetic
fields, which have combined the radio imaging of the large-scale radio structure of M\,87 with the VLA at 90\,cm, the B-fields at different sites of the lobe within a region about 40\,kpc in size are greater than or equal to $6.9\,\mu$G, see Table.\,I in Ref.\,\cite{Owen2000}. Hence, we conservatively assume a homogeneous B-field for the lobe with strength $5\,\mu$G, as shown in Table.\,\ref{table:1}.

 \emph{IC\,310}(Gal\,Long: 150.8, Gal\,Lat: -13.3; $z$=0.019)---seems to be a transitional AGN between a low-luminosity HBL (high-frequency
peaked BL Lac, namely blazar with weak optical
emission lines \cite{Urry1995}) and a radio galaxy \cite{Franceschini2019}, located on the outskirts of the Perseus galaxy cluster. An extraordinary TeV flare in 2012 Nov.\,12-13 and then a low state during several of the following months was detected by MAGIC \cite{IC310MAGIC2017}. A high state also has been observed by the MAGIC telescopes between 2009 October and
2010 February. The photon spectral index at VHE band is 2 during an effective observation of 3.5\,h in the (2009/2010) high state, and in this paper we adopt this high state observations together with the first three year (2008-2011) data taken with Fermi-LAT \cite{Aleksic:2013bya} (see Fig.\,\ref{fig:constrain}).

This head-tail radio galaxy has a radio lobe with a length of about 350\,kpc ($15^{\prime}$) \cite{Feretti:1997pz,Sato:2005mc}. The equipartition values of the lobe magnetic-field derived by Ref.\,\cite{Feretti:1997pz} (Table.\,2) decrease from 4.3\,$\mu$G at a distance of 10\,kpc to the core to 1.9\,$\mu$G at 140\,kpc. However, based on the observation at 610\,MHz, Fig.\,10 of Ref.\,\cite{Sebastian:2017wcd} shows that the equipartition
magnetic-field decrease from 3\,$\mu$G at a distance of about 1\,kpc to the head to 1.5\,$\mu$G at 70\,kpc. Besides, from an XMM-Newton observation of IC\,310, Ref.\,\cite{Sato:2005mc} derived the lower limit of the magnetic field strength to be $B>1\mu$G. Accordingly, we assume the strength of the lobe B-field is 2\,$\mu$G, as shown in Table.\,\ref{table:1}.

\emph{Mrk\,501}(Gal\,Long: 60.3, Gal\,Lat: 38.86; $z$=0.034)---the next-closest known HBL. It is known for showing the spectral variability at VHE band. During a long outburst observed with HEGRA in 1997, the source showed a very hard intrinsic spectrum with no softening up to the highest-energy detected photons of 20\,TeV \cite{Aharonian:2000xr, Franceschini2019}. We choose this spectrum that is detected during much of the 110\,h observation time, spread over 6 months \cite{Aharonian:2000xr} (see Fig.\,\ref{fig:constrain}). It is worth mentioning that, on the night of 2014 June 23-24, a flare comparable to the 1997 maximum was observed by HESS \cite{Abdalla:2019krx}.

There are no precise observations for the lobe of the BL Lac Mrk\,501. But BL Lacs are the aligned counterparts of FR\,I radio galaxies in the famous unified schemes for radio-loud AGN \cite{Urry1995}. Hence, we assume that the lobe B-field for this source is the same as IC\,310, which is classified into a transitional population between BL Lac objects and FR\,I radio galaxies \cite{Aleksic:2013bya}.
Furthermore, there is observational evidence that Mrk\,501 is also a member of a small cluster \cite{Stocke1995, Nilsson1999}. This source is believed to be located in galaxy clusters by Ref.\,\cite{hardening2} in the study of the photon-ALP mixing. Hence, it seems reasonable to assume $B_{\rm s}=$2\,$\mu$G.

For simply, we conservatively assume the region sizes of the lobe B-field for the three sources $r_{\rm s}=$10\,kpc in our fiducial scenario, as shown in Table.\,\ref{table:1}.

Finally, we model the Galaxy B-field with a homogeneous and isotropic field, however the realistic magnetic-field in the Milky Way depends on the source position in the sky and thus the back-conversion probability $P^{\rm G}_{a\gamma}$ for an ALP at the edge of the Galaxy to convert into a photon at Earth also depends on the source position \cite{hardening2}. Therefore we should ensure our sample at least are not in the position with relatively weak conversion. With the help of FIG.\,2 in Refs.\,\cite{hardening2}, which gives an illustrative sky map of the line-of-sight dependent probability for the common Jansson and Farrar magnetic field model \cite{Jansson2012,Fermi2016alp}, we find the three sources under consideration are just located in regions where the back-conversion probabilities are in the moderate intensity range. Also NGC\,1275 (NGC\,1275 and IC\,310 are both in the Perseus galaxy cluster) \cite{Fermi2016alp} and Mrk\,501 \cite{hardening2,Meyer2013,Galanti2019} have been studied to research ALPs involving photon-ALP conversion in the Galaxy B-field.

\subsection{Theoretical and intrinsic spectra}
\label{sec:Intr-spectra}
We model VHE gamma-ray spectra with
\begin{equation}\label{eq:model}
\psi_{0}=\rm e^{-\tau_{\gamma \gamma}}\phi\,\,\rm or\,\, \psi_{1}=P_{\gamma\rightarrow\gamma}\phi,
 \end{equation}
 where $P_{\gamma\rightarrow\gamma}$ and $\tau_{\gamma \gamma}$ are defined in Eq.\,\ref{PSP}. $\phi$ represents the intrinsic spectrum assumed for the sources. The model with ALP has two additional free parameters, $g_{a\gamma}$ and $m_{a}$, relative to the traditional model.

    $\phi$ is assumed as one of the three common models \cite{HESS2013}: power-law with exponential cut-off (PLC), log-parabola with exponential cut-off (LPC) and broken power-law with exponential cut-off (BPLC). The PLC spectrum is described by three parameters: $\phi_{\rm PLC}=\phi_{0}(E/E_{0})^{-\alpha}{\rm exp}(-E/E_{\rm c})$, where $E_{\rm c}$ is the cut-off energy, $\alpha$ is the photon spectral index constrained by the particle acceleration theory as $\alpha\geq1.5$, $\phi_{0}$ is the flux normalization, and $E_{0}$ is the fixed reference energy. While the LPC spectrum has additional curvature parameter $t>0$:
 $\phi_{\rm LPC}=\phi_{0}(E/E_{0})^{-s-t\, {\rm log}(E/E_{0})}{\rm exp}(-E/E_{\rm c})$ and also $\langle s+t\, {\rm log}(E/E_{0})\rangle \geq1.5$. The BPLC spectrum $\phi_{\rm BPLC}=\phi_{0}(E/E_{0})^{-\alpha_{1}}[1+(\frac{E}{E0})^{g}]^{\frac{\alpha_{1}-\alpha_{2}}{g}}{\rm exp}(-E/E_{\rm c})$ has 5 parameters.

  Since the EBL absorption-corrected spectrum of Mrk\,501 and even the other two observed ones in our samples are hard with no sign of convergence, the observations together with the chosen intrinsic spectrum only can give a relatively weak constraint on $E_{\rm c}$ and in principle give the minimum $E_{\rm c}$ above which the goodness of best-fit (parameterized by the minimum $\chi^2/\rm d.o.f.$) remains unchanged. Here, we take the minimum $E_{\rm c}$ as the fiducial value, see section.\,\ref{sec:result} and Table.\,\ref{table:1}. The highest energy of the detected photons in our samples is $\lesssim$ 20\,TeV, therefore the minimum $E_{\rm c}$ determined by the observations should above 20\,TeV. If the parent particles responsible for the VHE emission are electrons, the cutoff can be derived from the Klein-Nishina suppression, energy loss of the electrons and pair attenuation in the VHE emission region (see e.g., \cite{Lefa2012,Stawarz2008,Lewis2019,Warren2020,Lemoine2020,Mrk501fermi2011}), and the cutoff is more likely to be less than 100\,TeV. Hence, we uniformly take $E_{\rm c}$=100\,TeV as the optimistical one for our samples \cite{Franceschini2019}, even though $E_{\rm c}$ could be higher if the VHE $\gamma$-ray emission is of a hadronic origin \cite{LHAASO2019,Xue2019}.

 \subsection{Fitting and extrapolating the observations}
\label{sec:fitting}
To simulate the observations at the highest energies, we firstly fit the three observed spectra with $\psi_0$ and extrapolate it to hundreds of TeV energies, respectively. Meanwhile, the form (PLC, LPC or BPLC) of intrinsic spectrum $\phi$ to be chosen for each source is determined in terms of its minimum chi-square value per degree of freedom ($\chi^2/\rm d.o.f.$), and we find the minimum $E_{\rm c}$ above which the goodness of fit of $\psi_0$ with the best chosen $\phi$ remains unchanged. Then, we use $\psi_1$ containing the determined $\phi$ and $P_{\gamma \rightarrow \gamma}$ to fit and extrapolate the observations of each source, under the given fiducial $B_{\rm s}$, $r_{\rm s}$, $g_{a\gamma}$ and $m_{a}$.

 We repeat the process above with the optimistic parameters in Table.\,\ref{table:1} for the optimistic case. In this process, we take $E_{\rm c}=100$\,TeV for the chosen intrinsic spectrum $\phi$ above, but other parameters of $\phi$ such as the spectral index can change when fitting.

 We assume if the ALP-induced flux enhancement $\frac{\psi_1}{\psi_0}$ is more than one order of magnitude and the predicted spectra $\psi_{1}$ is over the equipment sensitivity, then the given ALP could be constrained. As the continuity and (approximative) monotonicity of $P_{\gamma \rightarrow \gamma}$, we only need to test the ALP parameters with small $g_{a\gamma}$ and that with large $m_a$ to obtain the constrained region.

 \begin{figure}[t]
\centering
\includegraphics[width=0.48\textwidth]{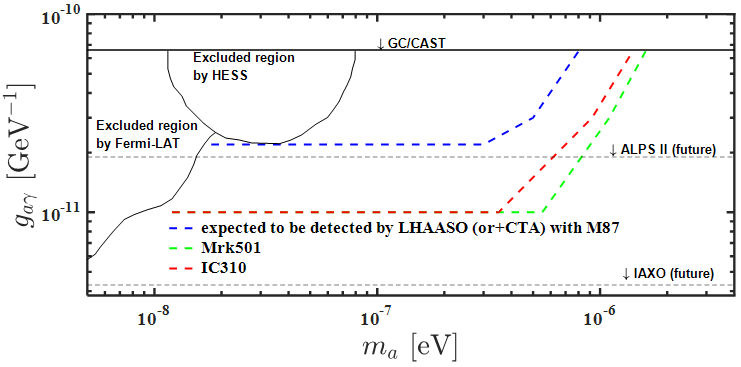}
\caption{\small{The expected ALP limits, based on our model as well as the future LHAASO (or+CTA) observations on the three sources for the optimistic parameters in Table.\,\ref{table:1}. For these limited parameters, the ALP-induced flux enhancement is more than an order of magnitude. The limit from Mrk\,501 and M\,87 are predicted with 5-years sensitivity of LHAASO and IC\,310 with one year. The meanings of various symbols in the diagram are the same as Fig.\,\ref{fig:constrain}.}}
\label{fig:constrain_opt}
\end{figure}

\section{Results}
\label{sec:result}
Crucial parameters for the best-fitting intrinsic spectra and the goodness of fit $\chi^{2}/$d.o.f. are listed in Table.\,\ref{table:1}.
Fig.\,\ref{fig:constrain}, Fig.\,\ref{fig:appendix} in the appendix and Fig.\,\ref{fig:constrain_opt} report the predicted sensitivity limits for future LHAASO/CTA observations of the three most promising nearby AGNs. Fig.\,\ref{fig:constrain} together with Fig.\,\ref{fig:appendix} correspond to the fiducial parameters in Table.\,\ref{table:1} and Fig.\,\ref{fig:constrain_opt} the optimistic ones. The blue line in Fig.\,\ref{fig:constrain} and Fig.\,\ref{fig:appendix} represents the standard absorption fit including an extrapolation above 100\,TeV. The results corresponding to the minimum allowable coupling $g_{a\gamma}$ for different mass $m_a$ are shown by the red line. The 50-h 5\,$\sigma$ sensitivity limits for CTA, 5-year and 1-year 5\,$\sigma$ limits for LHAASO are shown with the black dotted line, the black dashed line and the black solid line, respectively. Combined observations from the instruments will reach sensitivities of a few times $10^{-11}$\,GeV\,$\rm cm^{-1}$\,$\rm s^{-1}$.

\emph{M\,87}. The best-fitting intrinsic spectrum for the VHE observations is PLC with goodness of fit $\chi^2/$d.o.f.=1.2.
 and the minimum cutoff energy is $E_{\rm c}=90$\,TeV, see Fig.\,\ref{fig:constrain} and Fig.\,\ref{fig:appendix}. The predicted spectra extrapolated by the best-fitting model ($\psi_0$ and $\psi_1$) are above the 5-year sensitivity of LHAASO up to 70\,TeV (this will allow measurements
of the M\,87 spectrum up to about 70\,TeV), which is beneficial to constrain the intrinsic spectrum. Above 100\,TeV, the photon-ALP conversion gradually become important so that the photons survive mainly thought $\gamma \rightarrow a\rightarrow \gamma$ channel. Consequently, the flux enhancement is more than an order of magnitude and the flux is over the LHAASO sensitivity around 100\,TeV for three given ALP-parameter values. The line of $\psi_1$ take a very shallow ``valley'' or descending shape, as the low redshift of M\,87 lead to a smooth transition of survival probability $P_{\gamma\rightarrow\gamma}\approx P_{1}$ to $P_{\gamma\rightarrow\gamma}\approx P_{2}$. At the highest energy band, the intrinsic-spectrum cutoff makes the curve go down.

\emph{IC\,310}. The spectrum shows an obvious break at 200\,GeV and a BPLC model can fit the spectrum very well with goodness of fit $\chi^2/$d.o.f.=0.18, see Fig.\,\ref{fig:constrain} and Fig.\,\ref{fig:appendix}. The minimum cutoff energy is constrained to $E_{\rm c}=40$\,TeV. The predicted spectra extrapolated by the best-fitting model ($\psi_0$ and $\psi_1$) are above the sensitivity of LHAASO up to about 30\,TeV, which theoretically will allow measurements
of the IC\,310 spectrum up to 30\,TeV. Above $\sim$50\,TeV, the photon-ALP conversion gradually become important, so that the flux enhancement is over an order of magnitude and is above the one-year LHAASO sensitivity around about 100\,TeV. But the intrinsic-spectrum cutoff or it together with the CM effect makes the curve turn down at the highest energy band.

\emph{Mrk\,501}. The spectrum after EBL-absorption correction obviously takes the shape of a broken power-law \cite{Franceschini2019}, so the best-fitting intrinsic spectrum is BLPC with goodness of fit $\chi^2/$d.o.f.=1.4, see Fig.\,\ref{fig:constrain} and Fig.\,\ref{fig:appendix}. The minimum cutoff energy is $E_{\rm c}=70$\,TeV. The predicted observed spectrum without ALP is over the five-years sensitivity of LHAASO up to about 30\,TeV, which theoretically will give a relatively weak constraint on the intrinsic spectrum up to above 100\,TeV. The very prominent enhanced flux at around 100\,TeV is over the LHAASO sensitivity.

We estimate the ALP parameter space that will be possible probed by the future LHAASO (or+CTA) observations in the last picture of Fig.\,\ref{fig:constrain} for the fiducial scenario, where other limits and sensitivity projections are also given for comparing. For M\,87, LHAASO would be able to explore $g_{11}$ down to about 2.3 for $m_a<0.3\,\mu$eV. For IC\,310 and Mrk\,501, a lower value of $g_{11}\simeq 1.9$ for $m_a<0.2\,\mu$eV and $g_{11}\simeq 1$ for $m_a<0.2\,\mu$eV would be explored respectively, some of which is invoked to explain the cold dark matter \cite{Fermi2016alp}. The results corresponding to Mrk\,501 give the strongest exploitable bound on the coupling for $m_a \,\lesssim0.5\mu$eV: $g_{11}\simeq2$. In the case of M\,87, a relatively weak bound is given, as its observed flux is lower and its lower redshift leads to that higher energy is required to achieve the same enhancement.

For the optimistic scenario in Table.\,\ref{table:1}, the ALP parameter space that will be possible probed is shown as Fig.\,\ref{fig:constrain_opt}. No surprise, it predicts stronger restrictions than the fiducial scenario. For M\,87, the ALP parameters expected to be probed are almost unchanged compared to the fiducial scenario, though the source B-field increases by a factor 2 and the cutoff energy increase to 100\,TeV from 90\,TeV. In the case of IC\,310, the limited parameter space significantly expands to $g_{11}\gtrsim2$ for $m_a \,\lesssim0.35\mu$eV, benefiting from the increase of cutoff energy from 40\,TeV to 100\,TeV and increases of the source B-field by a factor 5. Though the parameter region from Mrk\,501 becomes wider owing to the increase of cutoff energy from 70\,TeV to 100\,TeV and increases (same as IC\,310), it is only slightly larger than that from IC\,310. Therefore, the result seems to more sensitive to the cutoff energy than the source B-field.

\section{Discussion}
\label{sec:discussion}
In this section, we will discuss our model assumptions about the source B-field, the cutoff energy of the intrinsic spectrum and the results from IC\,310.
\begin{figure}
\centering
\includegraphics[width=0.8\columnwidth]{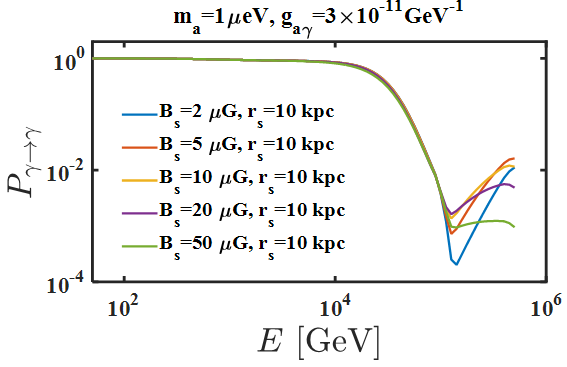}
\caption{\small{The photon survival probabilities $P_{\gamma\rightarrow\gamma}$ for different values of the source field magnetic $B_{\rm s}$. The redshift $z=0.005$ and field-magnetic region
$r_s$=10\,kpc. For $B_{\rm s}$ in the range from 5\,$\mu$G to 20\,$\mu$G (even if $B_{\rm s}\gtrsim20$\,TeV were too high and unrealistic for FR\,I lobe), the photon survival probabilities are close at around 200\,TeV. When $B_{\rm s}=2\,\mu$G, the critical energy of photon-ALP conversion is higher, and thus the ALP-induced enhancement occurs at higher energy. The CM effect almost completely suppresses the photon-ALP conversion for $B_{\rm s}=50\,\mu$G.}}
\label{fig:B}
\end{figure}

As one of the effective conversion conditions requires the photon energy to satisfy $E_{\rm crict}<E<E_{\rm H}$ and depends on the $B-$field, particularly the highest energy conversions in the source $B-$field is prone to be suppressed by the CM effect. Therefore, the uncertainty of $B_{\rm s}$ can translate into an uncertainty of the photon survival probability and our result. Fig.\,\ref{fig:B} shows how $B_{\rm s}$ affects $P_{\gamma\rightarrow\gamma}$ for a fixed redshift $z=0.005$ and ALP parameters $m_a=1\,\mu$eV, $g_{11}=3$. For $B_{\rm s}$ in the range from 5\,$\mu$G to 20\,$\mu$G (even if $B_{\rm s}\gtrsim20\,\mu$G were too high and unrealistic for FR\,I lobe), the photon survival probabilities are close at around 200\,TeV. It means that provided $B_{\rm s}$ is between 5\,$\mu$G and 20\,$\mu$G the ALP-induced flux enhancement could be achieved and comparable as done with $B_{\rm s}=10\,\mu$G above. When $B_{\rm s}=2\,\mu$G, the critical energy of photon-ALP conversion is close to 100\,TeV, and thus the ALP-induced enhancement occurs at higher energy. At the highest energy, $P_{\gamma\rightarrow\gamma}$ for $B_{\rm s}=2\,\mu$G can rise to that resulted from $B_{\rm s}=10\,\mu$G. When the mass $m_{a}$ is larger or the coupling $g_{a\gamma}$ is smaller and thus the critical energy is higher, $P_{\gamma\rightarrow\gamma}$ with $B_{\rm s}=10\,\mu$G can be about a factor five larger than $P_{\gamma\rightarrow\gamma}$ with $B_{\rm s}=2\,\mu$G, as shown in Fig.\,\ref{fig:B}.
\begin{figure*}[t]
\centering
\includegraphics[width=0.56\textwidth]{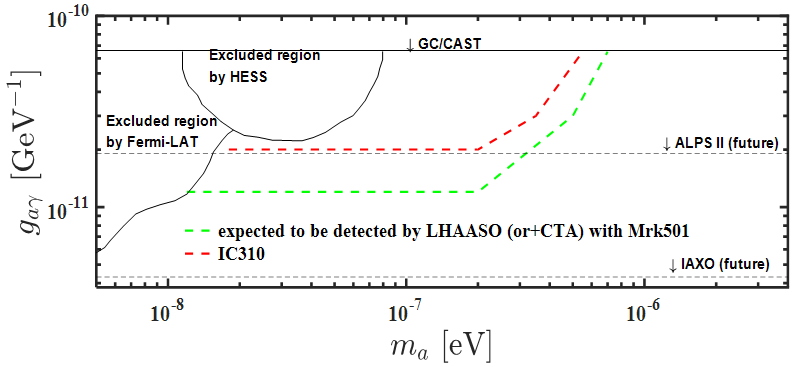}
\includegraphics[width=0.43\textwidth]{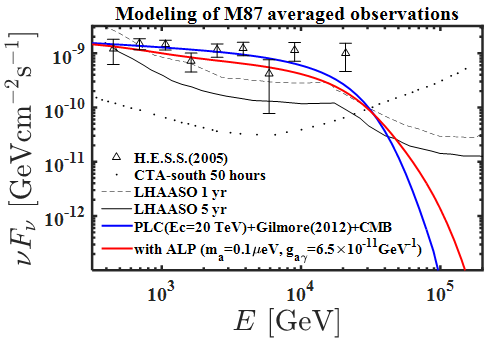}
\caption{\small{The results for the parameters in Table.\,\ref{table:2}, where the cutoff energy $E_{\rm c}$=20\,TeV for each spectrum. Right: expected ALP limits, based on our model as well as the future LHAASO (or+CTA) observations of IC\,310 (red dash line) and Mrk\,501 (green dash line). The limit from Mrk\,501 is predicted with 5-years sensitivity of LHAASO and IC\,310 with one year. For these limited parameters, the ALP-induced flux enhancement is more than an order of magnitude. Left: fitting and extrapolating the observations of M\,87. The detectable ALP-induced flux enhancement $\frac{\psi_1}{\psi_0}$ for unconstrained ALP parameters is less than 10 for M\,87, so we consider that this spectrum and our model can not give the limit on ALP. The meanings of various symbols in the diagram are the same as Fig.\,\ref{fig:constrain}.}}
\label{fig:constrain_20TeV}
\end{figure*}

But the intrinsic spectral exponential cutoff can lead to an exponential reduction of the source photons and thus indirectly reduce the photon-survival possibility with the same way. It means that our results are more sensitive to the uncertainty from the cutoff energy than the source B-field.

\begin{table}[t]
\centering
\caption{The lobe magnetic fields and intrinsic spectra with a certain cutoff energy $E_{\rm c}$=20\,TeV for every source. The values of the B-field parameters are fiducial as in Table.\,\ref{table:1}, while we take smaller cutoff energy $E_{\rm c}$=20\,TeV even though it is disfavour by the current limited observations, i.e.,\,the minimum $\chi^{2}$/d.o.f. of the $\psi_0$ (without ALP) fitting for each spectrum is significantly larger.
\label{table:2} }
\renewcommand\tabcolsep{0.1pt}
\begin{ruledtabular}
\begin{tabular}{lccccc}
   Source&$B_{\rm s}$\,($\mu$\,G) & $r_{\rm s}$\,(kpc)& $E_{\rm c}\,$(TeV)& $\alpha$\,(model) & $\chi^{2}$/d.o.f.\\
\hline
M\,87& 5 & 10 & 20 &
 2.1\,(PLC) & 1.4\\
IC\,310& 2 & 10 & 20 &
 1.18,1.74\,(BPLC) & 0.23 \\
Mrk\,501& 2 & 10 & 20 &
 1.83,2.28\,(BPLC) & 1.6 \\
\end{tabular}
\end{ruledtabular}
\end{table}

Our results are sensitive to the cutoff energy $E_{\rm c}$, as the predicted photons around 100\,TeV are mainly from the conversion channel $\gamma\rightarrow a\rightarrow \gamma$. In the fiducial scenario, the minimum cutoff energy is constrained by a limited amount of observations and the assumed intrinsic spectrum. The cutoff energy might still be more lower if it were obtained by fitting more comprehensive observations, especially those at higher energy. We therefore investigate the limit, where we assume the cutoff energy $E_{\rm c}$ is equal to the highest photon energy of 20\,TeV in our sample. Crucial parameters for the best-fitting intrinsic spectra and the goodness of fit $\chi^{2}/$d.o.f. are listed in Table.\,\ref{table:2}. The goodness of fit ($\chi^{2}/$d.o.f.) in this case is noticeably worse compared to the fiducial case. The B-field parameters are the same as the fiducial ones. We find the limited ALP-parameter region drastically reduces to $g_{11}\gtrsim 2$ with $m_{a}\lesssim0.3\,\mu$eV and M\,87 can not give any constraint due to the weak detectable ALP-induced enhancement, see Fig.\,\ref{fig:constrain_20TeV}.

We discuss the possible of $E_{\rm c}=100$\,TeV assumed in the optimistic case in two ways bellow.

Firstly, the optical depths due to internal pair-creation at such high energy can lead to an intrinsic spectral cutoff. In simple one-zone SSC scenarios, if requiring the optical depth $\tau$ less than 1, then the jet Doppler factor must satisfy (e.g. Refs.\,\cite{Mrk501fermi2011,Dondi1995})
\begin{equation}
\delta\gtrsim8.6(1+z)^{\frac{1}{3}}(\frac{E}{100\,\rm TeV}\frac{F_{0}}{10^{-11}\,\rm erg\,cm^{-2}\,s^{-1}}\frac{1\,\rm day}{t_{\rm var}})^{\frac{1}{6}},\label{Doppler}
\end{equation}
where $F_{0}$ is the observed monochromatic flux
energy density as measured at the observed photon energy $\epsilon_{0}$, $t_{\rm \,var}$ is an observed minimum (typical) variability time scale, and $\epsilon_{0}\simeq0.5\,(\frac{100\,\rm TeV}{E})(\frac{\delta}{10})^{2}$\,eV. Usually, the flux $F_{0}$ increases slowly with the photon energy $\epsilon_{0}$ and thus with $\delta$ up to the synchrotron peak, see e.g. Refs.\,\cite{Mrk501fermi2011}. Hence, we fix $F_{0}$ at near-infrared energy $\epsilon_{0}\simeq1$\,eV ($2.4\times10^{14}$\,Hz) for $E=100$\,TeV and $\delta\simeq14$. For Mrk\,501, $F_{0}\simeq5\times10^{-11}\,\rm erg\,cm^{-2}\,s^{-1}$ \cite{Petry2000} and $t_{\rm \,var}\simeq1$\,day \cite{Aharonian:1999vy}, so we derive $\delta\gtrsim11.1$ with  $E=100\,\rm TeV$ and Eq.\,\ref{Doppler}; Similarly, for M\,87,  $F_{0}\simeq3\times10^{-12}\,\rm erg\,cm^{-2}\,s^{-1}$ (derived from almost all of the infrared luminosity $L_{\rm\,IR}\simeq10^{41}\,\rm erg\,\rm s^{-1}$ of M\,87 observed in 2000 \cite{Whysong:2002ks}), $t_{\rm \,var}\simeq2$\,days \cite{M87HESS2006}, $\delta\gtrsim6.3$, and for IC\,310, $F_{0}\simeq1\times10^{-12}\,\rm erg\,cm^{-2}\,s^{-1}$ (inferred from the SSC model that is fitted the multi-frequency observations \cite{IC310MAGIC2017}), $t_{\rm \,var}\simeq0.55$\,day \cite{Aleksic:2013bya}, $\delta\gtrsim6.6$. These derived values of Doppler factor are not in tension with the typical values, especially for Mrk\,501 \cite{Mrk501fermi2011}. Note that even if $\tau>1$, the VHE flux would not reduce exponentially by exp($-\tau$), but only by a factor of $\tau$ (i.e. lead to a spectral break), owing to those photons produced in the last transparent layer are still able to escape unabsorbed \cite{Rieger:2007tt,Neronov:2007vy}. If the intrinsic absorption is outside the VHE region along line of sight, the cutoff is exponential, but it is complicated to estimate it and may be less important \cite{Aleksic:2013bya}.

  Secondly, even though the spectrum is dominated by emission of leptonic origin (with evidence that most of the rapid variable emission has a leptonic origin), the emitted spectrum with cutoff above 100\,TeV at VHE region is still possible. The recent observation from the Crab Nebula with energy beyond 100\,TeV show no exponential cutoff below 100\,TeV, which is usually interpreted in the framework of leptonic models \cite{Amenomori2019,MAGIC2019,HAWC2019}. As powerful cosmic particle accelerators \cite{Kotera2011}, that may happen on some extreme TeV AGNs, too. Furthermore, AGNs are excellent candidates as Ultra-High-Energy Cosmic Rays sources \cite{Mbarek2019}, and the hadronic cosmic rays are capable of producing spectrum without sharp cutoff below 100\,TeV if the VHE emission is dominated by hadronic origin \cite{CTA2017}.

  To determine the magnitude of $E_{\rm c}$ without ambiguity, we need to further research on the intrinsic physics (including parent particle species and its spectral energy distribution, the radiation mechanism, and pair attenuation in the emission region) of the $\gamma-$ray sources, as well as the forthcoming observations above tens of TeV by CTA, LHAASO, SWGO and so on.

  The VHE emission from IC\,310 only have been detected two times by MAGIC so far \cite{MAGIC2019}, which are relatively sparse compared to the two other sources. Therefore, we are more strict with IC\,310 and simulate its observations at high state between 2009 October and 2010 February with 1\,yr sensitivities of LHAASO. If such high state would occur more than three times during the whole performance period of LHAASO, our expected constraint on ALPs is possible.

\section{Conclusion}
\label{sec:conclusion}
In this article, we have discussed the potential of the gamma-ray spectrum of AGN for energy up to above 100\,TeV to probe ALP parameter space at around $\mu$eV, where the coupling $g_{a\gamma}$ is so far relatively weak constraint.

In case of conventional physics, most of the photons above tens of TeV emitted from distant (distance$>$10\,Mpc) AGN would be absorbed by the EBL/CMB during its travel to the earth (see Fig.\,\ref{fig:1} and \ref{fig:pz}). But more such photons, no matter how far away, could survive, if we assume that the photon-ALP conversions ($\gamma\rightarrow a\rightarrow\gamma$) take place separately in the homogeneous lobe (or plume) and Galaxy magnetic-field. Consequently, a very significant ALP-induced flux enhancement, shaped as a peak, is expected to arise in the observed spectrum above tens of TeV (see Fig.\,\ref{fig:pz}). This provides the upcoming LHAASO a good chance to detect the enhancement as its unprecedented sensitivity above 30\,TeV.

In order to acquire as many observations at tens of TeV as possible and thus reduce the uncertainty from the intrinsic spectrum, the nearby and bright sources, such as Mrk\,501, IC\,310 and M\,87, are recommended to constrain the ALPs around $\mu$eV. Assuming an intrinsic spectrum with exponential cutoff, we have extrapolated the observed spectra of our sample up to above 100\,TeV by the models with/without ALPs. For $g_{a\gamma}\gtrsim 2\times$$10^{-11} \rm GeV^{-1}$ with  $m_{a}\lesssim0.5\,\mu$eV, the flux at around 100\,TeV predicted by the ALP model can be more than an order of magnitude larger than that from the standard absorption, and the enhanced flux could be detected by LHAASO (see Fig.\,\ref{fig:constrain} and \ref{fig:appendix}).

Our result is subject to the uncertainty from the extrapolation of intrinsic spectrum above tens of TeV and the source magnetic-field (see Fig.\,\ref{fig:constrain_opt} and \ref{fig:constrain_20TeV}). For an optimistic estimation, the constraint can be improved to $g_{a\gamma}\gtrsim 2\times$$10^{-11} \rm GeV^{-1}$ with $m_{a}\lesssim1\,\mu$eV (see Fig.\,\ref{fig:constrain_opt}). This will require further research on these sources (Mrk\,501, IC\,310 and M\,87) based on the forthcoming observations by CTA, LHAASO, SWGO and so on.

\begin{acknowledgements}
We would like to thank Weipeng Lin, P. H. T. Tam, Chengfeng Cai, Zhao-Huan Yu, Seishi Enomoto, Yi-Lei Tang and Yu-Pan Zeng for useful discussions and comments.
This work is supported by the National Natural Science Foundation of China (NSFC) under Grant No. 11875327, the Fundamental Research Funds for the Central Universities, China, and the Sun Yat-Sen University Science Foundation.
\end{acknowledgements}

\clearpage

\widetext
\appendix
\section{fitting results for the other different ALP parameters}
\label{appendix}
As a supplement of Fig\,\ref{fig:constrain}, we show the fitting results for another different bounded ALP parameters, where the observations of M\,87, IC\,310 and Mrk\,501 are fitted and extrapolated for the case of the fiducial parameters in Table.\,\ref{table:1}. The meanings of various symbols in the diagram are the same as Fig.\,\ref{fig:constrain}.
\begin{figure*}[b]
\centering
\includegraphics[width=0.45\textwidth]{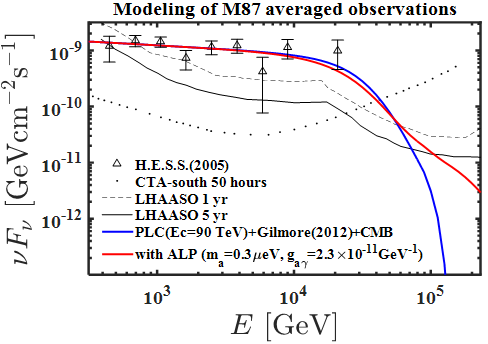}
\includegraphics[width=0.45\textwidth]{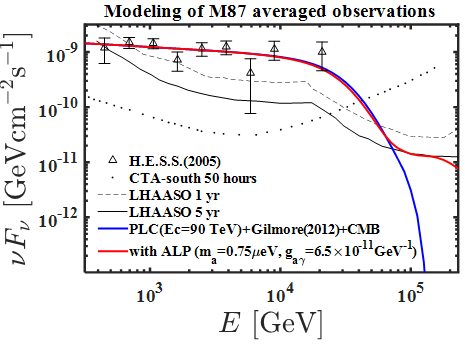}

\includegraphics[width=0.45\textwidth]{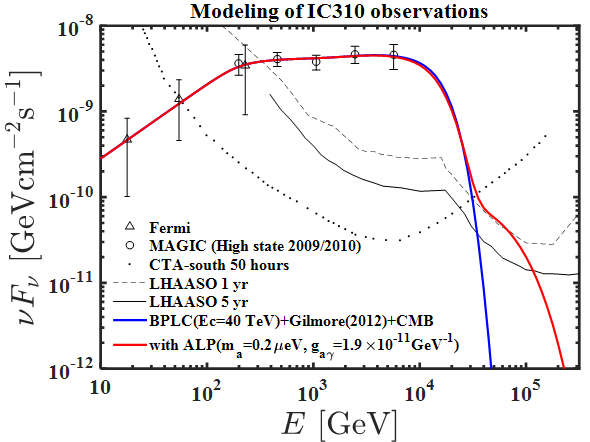}
\includegraphics[width=0.45\textwidth]{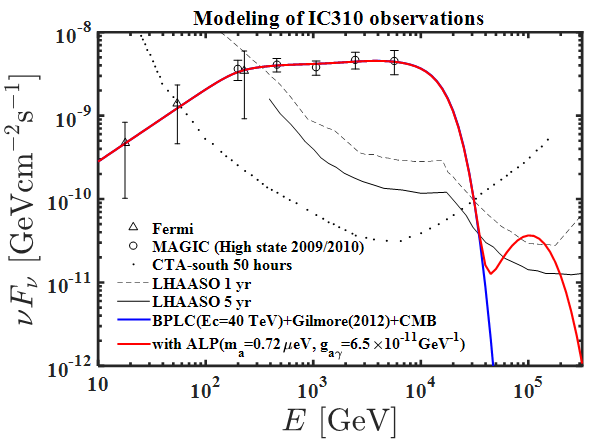}

\includegraphics[width=0.45\textwidth]{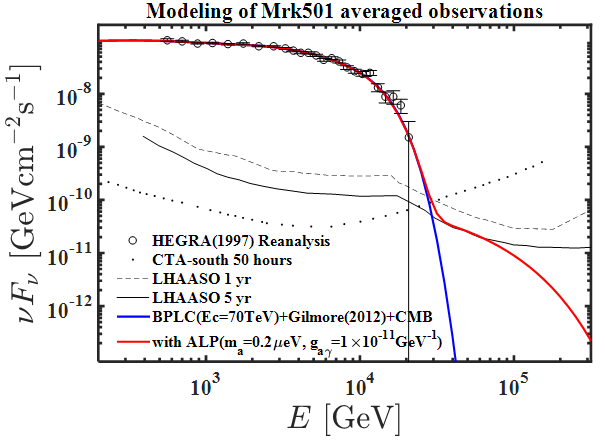}
\includegraphics[width=0.45\textwidth]{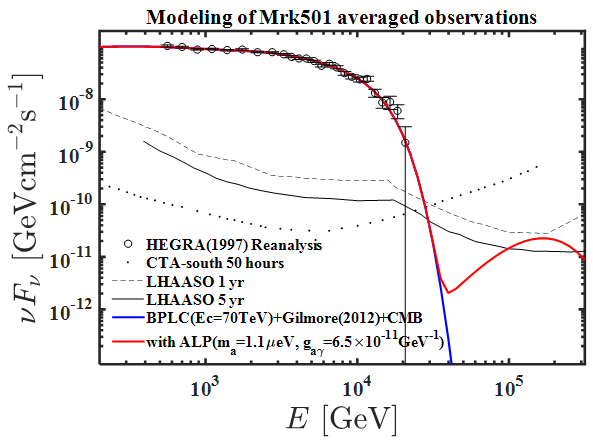}
\caption{The results for the fiducial parameters in Table.\,\ref{table:1}. The meanings of various symbols in the diagram are the same as the top panel and left of the bottom panels of Fig.\,\ref{fig:constrain}.}
\label{fig:appendix}
\end{figure*}

\clearpage

\end{document}